%Paper: comp-gas/9408001
%From: lopez@roxanne.nuclecu.unam.mx (Ramon Lopez)
%Date: Fri, 26 Aug 1994 16:16:48 -0600

\magnification=1200
\vsize=21.0 truecm
\hsize=14.5 truecm
% \hsize=16.0 truecm
\hoffset=4.5 truecm
% \hoffset=3.0 truecm
\voffset=3.5 truecm
% lo que pide la revista \baselineskip=22pt
\baselineskip=12pt

\font\ftit=cmbx10

\parskip=6pt
\parindent=2pc

\font\titulo=cmbx10 scaled\magstep1

\font\formula=cmti10 scaled\magstep2

\def\be{\hbox{\formula e}}
\def\section#1{\vskip 1.5truepc plus 0.1truepc minus 0.1truepc
	\goodbreak \leftline{\titulo#1} \nobreak \vskip 0.1truepc
	\indent}
\def\frc#1#2{\leavevmode\kern.1em
	\raise.5ex\hbox{\the\scriptfont0 $ #1 $}\kern-.1em
	/\kern-.15em\lower.25ex\hbox{\the\scriptfont0 $ #2 $}}

%The following command places a small circle over a character:

  %"equal by definition"
   %short for "one half"

%The following defines the character for Lie derivative. The quantity
%with respect to which the Lie derivative is taken should follow the
%command "\Lie", and is the one-parameter on which the def. depends:

   %semi-direct product
  %for real numbers
 %for complex numbers \def\covD{{\rmI\!D}}

%for covariant derivative "D" \def\Dirac{D\!\!\!\!/}
%for Dirac's operator "slashed D"
\newbox\pmbbox
\def\pmb#1{{\setbox\pmbbox=\hbox{$#1$}%
\copy\pmbbox\kern-\wd\pmbbox\kern.3pt\raise.3pt\copy\pmbbox\kern-\wd\pmbbox
\kern.3pt\box\pmbbox}}

% Hay que usar lo que sigue a continuacion para hacer el caracter
% bold \pmb4

\centerline{\ftit Recognition of Temporal Sequences of Patterns}

\centerline{{\ftit Using State-dependent Synapses}\footnote{*}{This
work is supported in part by CONACyT grant 400349-5-1714E and by the
Association G\'en\'erale pour la Coop\'eration et le D\'eveloppement
\break (Belgium).}}

\vskip 0.5pc

\centerline{F. Zertuche, R. L\'opez-Pe\~na${}^{\dagger}$ and H.
Waelbroeck${}^{\dagger}$}

\vskip 0.5pc

\centerline{Instituto de Investigaciones en
Matem\'aticas Aplicadas} \centerline{y en Sistemas, UNAM,
Secci\'on Cuernavaca, A.P. 139-B} \centerline{62191 Cuernavaca,
Morelos, M\'exico}
\centerline{e-mail: zertuche@ce.ifisicam.unam.mx}

\vskip 1.0pc

\centerline{${}^{\dagger}$Instituto de Ciencias Nucleares, UNAM}
\centerline{Apdo. Postal 70-543, M\'exico, D.F., 04510 M\'exico.}

\vskip 1.5pc

{\leftskip=3.5pc\rightskip=3.5pc\smallskip\noindent Short title:
Recognition of Temporal Sequences \smallskip}

\vskip 0.2pc

{\leftskip=3.5pc\rightskip=3.5pc\smallskip\noindent PACS numbers:
87.10.+e, 05.20.-y \smallskip}

\baselineskip=22pt

\vskip 1.5pc

\centerline { Abstract}

{\leftskip=1.5pc\rightskip=1.5pc\smallskip\noindent Using an
asymmetric associative network with synchronous updating, it is
possible to recall a sequence of patterns.  To obtain a stable
sequence generation with a large storage capacity, we introduce a
threshold that eliminates the contribution of weakly correlated
patterns.  For this system we find a set of evolution equations for
the overlaps of the states with the patterns to be recognized. We
solve these equations in the limit of the stationary cycle, and
obtain the critical value of the capacity as a function of the
threshold and temperature. Finally, a numerical simulation is made,
confirming the theoretical results.
	\smallskip}

\vfill
\eject

\section{1. Introduction}

The application of neural networks to time-sequence recognition has a
wide range of potential applications, from associative memory to
speech recognition to the prediction and control of complex dynamical
systems.  Networks with symmetric synapses, such as the
Hopfield~${}^{[1]}$ and Little~${}^{[2]}$ models are unable to
reproduce complex sequences since they minimize an energy
function~${}^{[3, 4]}$.  The equilibrium properties of these models
have been studied analytically by constructing the partition function
associated to the energy. In Ref.~[5] the asynchronous case (the
Hopfield model) and the synchronous one (the Little model) have been
studied far from saturation. That is to say, for $ \alpha = 0 $ with $
\alpha = p / N $, where $ p $ is the number of patterns and $ N $ is
the number of neurons as $ N $ goes to infinity. The near saturation
case ($ \alpha \not= 0 $) was studied by Amit {\it et al.}  for the
asynchronous case~${}^{[6]}$ and by Fontanari {\it et al.}  for the
synchronous one~${}^{[7]}$ by using replica symmetry techniques.
Networks with asymmetric synapses are able to reproduce sequences of
patterns, however in general the equilibrium statistical mechanics
cannot be applied. Instead of using an energy approach Coolen and
Ruijgrok~${}^{[8]}$ started from the master equation for the
microscopic states of the network and derived an evolution equation
for the probability density for the overlaps of the states with the
patterns in an asynchronous asymmetric network. Using this approach
other authors have studied asynchronous asymmetric nets far from
saturation~${}^{[9, 10]}$. The synchronous case was also analyzed far
from saturation by Bernier~${}^{[11]}$.

 Near saturation networks have been studied recently with success using
statistical methods based on the central limit theorem
(CLT). In Ref.~[12] Nishimori and Ozeki investigate the relaxation to
equilibrium in the Little model~${}^{[2]}$ (synchronous dynamics) near
saturation using a mean field treatment and solving the equations by
means of the CLT. They obtained in the equilibrium limit a good
qualitative agreement with the results by Amit Gutfreund and
Sompolinsky~${}^{[6]}$. An energy approach and the CLT was used by
Shukla ~${}^{[13]}$ in the Little model to obtain a phase diagram that
coincide well with the numerical results.  In Ref.~[14] we studied a
symmetric, asynchronous, state-dependent network with a threshold
using a CLT method developed by Geszti and Peretto~${}^{[15]}$ and
find that its capacity increases with $\eta $ (the threshold) as
$\alpha \sim \eta^{-1} \be^{\eta^2 / 2} $. Moreover for $ \eta = 0 $
we recovered the Hopfield model an found the same results as with
replica symmetry techniques~${}^{[6]}$.

   In this article, we will consider an asymmetric neural
network \break synchronously updated with a state-dependent synapses
and calculate its capacity to follow a sequence of patterns. For this
we first derive the evolution equations for the overlaps using the
transition probability for the neuron states and a mean field type
approximation.  We then solve them by the statistical method of Geszti
and Peretto.  We will show that this network has a maximal storage
capacity $ \alpha_c = p/N = 0.278 $ in the case $\eta = 0 $, which is
the limit in which the network is state-independent. Beyond this
value, the number of weakly correlated patterns becomes so large that
their contribution dominates over that of the highly correlated
pattern, which should force the transition to the next state in the
sequence, and the system diverges from the learned sequence.  We will
show how the storage capacity increases if one limits the
contribution of weakly correlated patterns, by taking $ \eta \not= 0
$: Only those patterns whose correlation with the state of the system
is greater or equal to the threshold are left to give a contribution
to the synapses.  We will give the dynamical phase diagram of this
network and compute the critical capacity as a function of the
temperature and the threshold: the learned sequence is stable up to
the critical capacity, which increases with the threshold. These
results are confirmed in the deterministic limit $ T = 0 $, with the
help of a numerical simulation.

Although biological synapses are likely to be asymmetric, and this
fact is probably an important factor in the complexity observed in
biological neural systems, the application of a state-dependent
threshold and the use of a central clock for the parallel updating of
the neurons implies that the results of this article are not likely
to be relevant in the physiological domain.  On the other hand, the
use of synchronous parallel updating allows for an efficient use of
modern parallel-processing computers. We will discuss possible
applications of our work to the prediction of chaotic time series in
the conclusion.

In Sec.~2, we review the increase in capacity with the threshold
parameter in the Hopfield model, as well as the solutions of the mean
field equations for symmetric synapses~${}^{[14]}$. In Sec.~3, we
consider the asymmetric rule with a threshold, and parallel updating
of the neurons; the evolution equations for the overlaps are derived
and solved using the statistical method developed in Ref.~[15], and
the generalized phase diagram for the stationary limit of the
dynamics is constructed. In Sec.~4, we give the results of the
numerical simulation and in Sec.~5 we give the conclusions.

\vfill
\eject

\section{2. State-Dependent Synapses and Associative Memory.}

In Ref.~[14] we considered a network with symmetric
state-dependent synapses for associative memory tasks. The main idea
of this type of synapses is to cut off the contribution of those
patterns whose correlation is less than a given value (the
threshold).  A set of $ N $ neurons with states $ s_i^n = \pm 1 $ at
time $ n = 1, 2,... $ ($i=1,...,N$) interact through a
state-dependent synapses matrix $ W_{ij}^n $.  The network evolves by
updating one neuron at a time in random order through the rule
	$$ s_i^{n + 1} = \pm 1 \qquad {\rm with \ probability}
	\qquad {1 \over 1 + \be^{\mp 2 \beta h_i^n}} \eqno(1) $$
	where
	$$ h_i^n = \sum_{j = 1}^N W_{ij}^n s_j^n  \eqno(2) $$
and where $\beta^{-1} = T$ is the temperature parameter. Let $\{
\xi^\mu_i = \pm 1 \}$ ($\mu = 1,...,p$) be a set of $ p $ patterns
generated randomly, such that
	$$ \ll \xi_i^\mu \xi_j^\nu \gg \ = \
	\delta^{\mu \nu} \delta_{i j}, \eqno(3) $$
	where $\ll ... \gg$ denotes the average over the distribution
of patterns. The synapses matrix was taken to be
	$$ W_{ij}^n = {1 \over N} \sum_{\mu = 1}^p \xi_i^\mu
	\xi_j^\mu \
	\Theta \left( \left(m_\mu \left( s^n \right) \right)^2 -
	{\eta^2 \over N}\right). \eqno(4) $$
	where
	$$ m_\mu \left( s^n \right) \equiv {1 \over N} \sum_{j = 1}^N
	\xi_j^\mu s_j^n \eqno(5) $$
	is the correlation of the state $ s_i^n $ with the pattern $
\xi_i^\mu $, and the step function $ \Theta \left( x \right) $
ensures that the pattern $ \xi^\mu $ is set to zero if $ \left( m_\mu
\left( s^n \right) \right)^2 < \eta^2 / N$. $ \eta $ is called the
threshold and for $ \eta = 0 $ Hebb's rule for the synapses is
recovered~${}^{[3, 11]}$. From (1) and (5) one obtains the equation
of motion for the instantaneous average activity of the neurons
	$$ m_\mu^{n + 1} \ = {1 \over N} \sum_j \xi_j^\mu \tanh
	\left\{ \beta \sum_\nu \xi_j^\nu m_\nu \left( s^n \right) \
	\Theta \left( \left( m_\nu \left( s^n \right) \right)^2 -
	{\eta^2 \over N} \right) \right\}, \eqno(6) $$
	where
	$$ m_\mu^n \ \equiv \ < m_\mu \left( s^n \right) > \ = \ {1
	\over N} \sum_{j = 1}^N \xi_j^\mu < s_j^n >, \eqno(7) $$
	$ < ... > $ denoting the thermal average over the
distribution (1). Since \break $ m_\mu^n - m_\mu \left( s^n \right)
\simeq O \left( {1 \over \sqrt{N}} \right) $ one can substitute $
m_\mu \left( s^n \right) $ by $ m_\mu^n $ in (6) making an error of
	$$ O \left( \sqrt{\alpha} \ Prob \left[ \left( m_\mu^n
	\right)^2 - {\eta^2 \over N} \geq 0 \right] \right)
	\eqno(8)$$
where $ \alpha = p / N $ and $ Prob \left[ x \geq 0 \right] $ is the
probability that $ x $ be greater or equal to zero.  In the limit $
\alpha \rightarrow 0 $ this substitution is exact, while for $ \alpha
\not= 0 $ the approximation remains good for large values of $ \eta $
since then the probability that a weakly correlated pattern pass the
threshold becomes small. With this approximation, one obtains the
mean field equations
	$$ m_\mu^{n + 1} \ = {1 \over N} \sum_j \xi_j^\mu \tanh
	\left\{ \beta \sum_\nu \xi_j^\nu m_\nu^n \Theta \left(
	\left(m_\nu^n\right)^2 - {\eta^2 \over N} \right) \right\}.
	\eqno(9) $$
	A study of these equations by means of the statistical
techniques developed in Ref.~[15] showed that the storage capacity $
\alpha_c = p/N$ can be increased up to any preassigned value (for $ N
\rightarrow \infty $) by taking a sufficiently large value of the
threshold. Those results were confirmed by means of a numerical
simulation.

\vskip 2.5pc

\section{3. Sequence Recognition with State-Dependent Synapses.}

   Let us consider a neural network with $ N $ neurons and a time
series of $q $ random patterns $ \{ \xi_i^\mu \} $ which obeys (3).
We want the network to be capable of following the sequence of
patterns in order from $ \mu = 1 $ to $ \mu = p $, where $ p < q $.
The capacity of the network is again defined by $ \alpha = p / N $.
The system evolves now in parallel by the transition probability from
the state $ {\bf s'} = \left( s'_i \right) $ to the state $ {\bf s} =
\left( s_i \right)$
	$$ \phi \left( {\bf s'} \to {\bf s} \right) \equiv \prod_{i
	= 1}^N \ {1 \over 2} \left[ 1 + s_i \tanh \left(\beta h'_i
	\right) \right] \eqno(10) $$
	where $ h'_i $ is given by (2) and the synapses is now
	$$ W_{ij}^n = {1 \over N} \sum_{\mu = 1}^p \xi_i^{\mu + 1}
	\xi_j^\mu \ \Theta \left( \left( m_\mu \left( s^n \right)
	\right)^2 - {\eta^2 \over N} \right). \eqno(11) $$
	The evolution of the probability of the network $ P_n \left(
{\bf s} \right) $ to be in state $ {\bf s} $ at time $ n $ is given
by~${}^{[10, 11]}$
	$$ P_{n + 1} \left( {\bf s} \right) = \sum_{{\bf s'}} \phi
	\left( {\bf s'} \to {\bf s} \right) P_n \left( {\bf s'}
	\right) \eqno(12) $$
The average overlap at step $ n + 1 $ is then
	$$ {\bf m}^{n + 1} \ = \ {1 \over N} \left< \sum_i \pmb\xi_i
			s_{i}^{n + 1} \right>_{n+1} \ , \eqno(13) $$
where
	$$ \left< A \right>_{n} \equiv \sum_{\bf s} P_{n} ( {\bf s} )
		\, A ( {\bf s} ) \ . \eqno(14) $$
After substituting (10) and (12) in the expression (13) and then
summing over {\bf s}, one finds
	$$ {\bf m}^{n + 1} \ = \ {1 \over N} \left< \sum_i \pmb\xi_i
	\tanh \left\{ \beta \sum_\mu \xi_i^{\mu + 1} m_\mu \Theta
	\left( \left( m_\mu \right)^2 - \eta^2 / N \right) \right\}
	\right>_n \eqno(15) $$
	We can substitute the overlaps in the right hand side of (15)
by its thermal average making an error of order given by (8), obtaining
	$$ m_\mu^{n + 1} \ = {1 \over N} \sum_j \xi_j^\mu
	\tanh \left\{ \beta \sum_\nu \xi_j^{\nu + 1} m_\nu^n \Theta
	\left( \left( m_\nu^n \right)^2 - {\eta^2 \over N} \right)
	\right\}. \eqno(16) $$
	We are interested in finding solutions of (16) which have
a large overlap only with the pattern $\mu = n$:
    	$$ \ll m_\mu^n \gg = \delta_\mu^n \ m_n. \eqno(17) $$
	Let us consider in (16) $ \mu = \nu + 1 \not= n + 1 $ and
expand the term proportional to $ m_\mu^n \Theta \left( \left(
m_\mu^n \right)^2 - \eta^2 / N \right) $ to first order. One finds
	$$ \eqalignno{ m_{\nu + 1}^{n + 1} &= {1 \over N} \sum_i
	\xi_i^{\nu + 1} \xi_i^{n + 1} \ \tanh \left[ \beta \left( m_n
	+ \Upsilon_i^{n, \nu} \right) \right] &\cr & \ + {\beta \over
	N} m_\nu^n \Theta \left( \left( m_\nu^n \right)^2 - {\eta^2
	\over N} \right) \sum_i \left[ 1 - \tanh^2 \beta \left( m_n +
	\Upsilon_i^{n, \nu} \right) \right] & (18) \cr} $$
where
	$$ \Upsilon_i^{n, \nu} \equiv \sum_{\mu \not= n, \nu}
	\xi_i^{\mu + 1} \xi_i^{n + 1} m_\mu^n \ \Theta \left( \left(
	m_\mu^n \right)^2 - {\eta^2 \over N} \right). \eqno(19) $$
Because the $ m_\mu^n $, $ \mu \not= n $ are the sum of a large number
of random variables \break $ \xi_i^\mu < s_i^n > $ we may assume that
they have normal distributions centered at zero, with variance
$\sigma_n^2 / N $, by the CLT.  For $ \eta \neq 0 $,
only the patterns which pass the threshold contribute to (19): the
distribution of $ m_{\mu}^{n} $ over the reduced set of patterns is
gaussian for $ ( m_{\mu}^{n} )^{2} > \eta^2 / N $ and zero in the
strip $ ( m_{\mu}^{n} )^{2} < \eta^2 / N $.  We will assume that the
number of patterns which pass the threshold is large, so that $
\Upsilon_{i}^{n, \nu} $ has a normal distribution also for $ \eta \neq 0
$, with variance $ \alpha r_{n} $ and average zero (note, however, that
the $ m_\mu^n $, $ \mu \not= n $ are not strictly independent
variables, since they are related through the equations (16)).  Then

	$$ \alpha r_n = \ \ll \left( \Upsilon_i^{n, \nu} \right)^2
	\gg, \eqno(20) $$
and
	$$ \eqalign{ q_n &\equiv {1 \over N} \sum_i \tanh^2 \beta
	\left( m_n + \Upsilon_i^{n, \nu} \right) \cr &= \int {dz
	\over \sqrt{2 \pi}} \be^{- z^2 / 2} \tanh^2 \beta \left( m_n
	+ \sqrt{\alpha r_n} z \right).\cr} \eqno(21) $$
	From (18) we get
	$$ \eqalign{ m_{\nu + 1}^{n + 1} - \beta m_\nu^n \Theta
	\left( \left( m_\nu^n \right) - {\eta^2 \over N} \right)
	& \left( 1 - q_n \right)  \cr &= {1 \over N} \sum_i
	\xi_i^{\nu + 1} \xi_i^{n + 1} \tanh \beta \left( m_n +
	\Upsilon_i^{n, \nu} \right) \cr} \eqno(22) $$
	Squaring this expression, averaging over the distribution of
patterns and using (19), (20) and (22) we get for $ p \gg 1 $
	$$ \sigma^2_{n + 1} - \beta^2 \left( 1 - q_n \right)^2 r_n =
	q_n \eqno(23) $$
Another relation can be obtained directly from equations (19)
and (20), giving
	$$ r_n = {2 \over \sqrt{\pi}} \sigma^2_n \ \Gamma \left( 3/2,
	{\eta^2 \over 2 \sigma^2_n} \right) \eqno(24) $$
	where $\Gamma$ is the incomplete gamma function${}^{[16]}$.
Taking now the mean field equations for $ \mu = n + 1 $ we obtain,
with similar calculations,
	$$ m_{n + 1} = \int {dz \over \sqrt{2 \pi}} \ \be^{- z^2 / 2}
	\ \tanh \beta \left( m_n + \sqrt{\alpha r_n} z \right).
	\eqno(25) $$

Equations (21), (23), (24) and (25) are simultaneous equations for
the unknowns $ m_n $, $ q_n $, $ \sigma_n $ and $ r_n $. We can solve
them in the stationary limit in which the order parameters are
constant in time: $ m_n = m $, $ q_n = q $, $
\sigma_n = \sigma $ and $ r_n = r $. In [Fig.~1], we describe
the space of solutions at $ T = 0 $: for $ \eta = 0 $ there
exist solutions for $\alpha \leq 0.278 $, while for
$ \eta \not= 0 $ this value increases
as is shown by the critical line $ \alpha_c \left( \eta \right) $; $
m \not= 0 $. Above this line one has temporal sequence
solutions of the form (11), while below it $ m = 0 $. For
$ \eta \to \infty $, $ \alpha_c \to \infty $ as
	$$ \alpha_{c} \approx \sqrt{{2 \over \pi}}
	\eta^{-1} \be^{{\eta^2 \over 2}} \ . $$

In [Fig.~2] and [Fig.~3] we give the critical capacity as a function
of the threshold for finite temperatures.

\vskip 2.5pc

\section{4. Numerical simulation}

A numerical simulation of a network of N =144 neurons was carried
out.  The network was initiated near the first pattern in each set,
with an error in one of the neurons, and allowed to evolve according
to the deterministic synchronous evolution rule with the synapses
(11).  The overlap with the last pattern in the sequence was
determined, and this result was averaged over $200$ sets of randomly
generated patterns and over 25 different choices of the erroneous
initial neuron.  For low values of the threshold the overlap begins to
decline sharply at a critical value of the capacity close to the
theoretical estimate.  For $ \eta \ge 2 $ the decline is smoother and
it is difficult to determine accurately the critical capacity.
Following the first referee's suggestion, we increased $N$ to 1681
neurons, using the ``trick'' described by Penna and Oliveira~${}^{[17]}$
(1681 is the square of 41; the odd number avoids memory drift
conflicts in the vectorized code).  The low-$ \eta $ graphs were
unchanged, but the new $ \eta = 2 $ graph displays a sharp drop near
the theoretical critical capacity estimate, namely at $ \alpha_c
\approx 1.1 $ [Fig.~ 5].

The need for a greater number of neurons for large-$ \eta $
simulations reflects the fact that the number of patterns which pass
the threshold must be large, in order to apply the CLT in (16).
Since $m_{\mu}$ has a normal distribution with variance $ {\sigma^2
\over N} \sim {1 \over N} $, this number is of the order $ \alpha N
\left( 1 - {\rm erf} \left( \eta / \sqrt{2} \right) \right) $, where
$ { \rm erf} () $ is the error function~${}^{[16]}$. For $ N = 144 $,
$ \eta = 2 $ and $ \alpha = 1 $, an average of only $ 6.55 $ patterns
pass the threshold, while for $ N = 1681 $ this number increases to $
76.5 $, enough to invoke the CLT.  As this interpretation suggests,
there should be no need to increase further the number of neurons;
this fact was confirmed by a run at $N = 6561$, which reproduced the
same graph as Fig.~5.

\vfill
\eject

\section{5. Conclusions}

We have proposed a neural network with state-dependent
synapses and synchronous updating capable of storing long sequences
of random patterns. In order to achieve this result, we introduced a
threshold which cuts off the contribution of the weakly correlated
patterns. The critical storage capacity was computed as a function of
the threshold parameter and the temperature, and the results for the
deterministic case $T=0$ were confirmed by numerical simulation.

One of the motivations for this work was the possibility of applying
this type of network to chaotic time series prediction.  There it is
necessary to ``teach'' the network a large number of patterns,
typically of the order of $10^4$ for a low-dimensional attractor.
Since the present network is capable of storing such a chaotic time
series, it is possible that as a dynamical system this network would
itself display a chaotic behavior, with an attractor which
approximates reasonably well the attractor of the physical system
that generated the time series.  Finally, it may be possible to
adjust the value of the threshold so that the first Liapunov
coefficient of the neural network be equal to that of the time
series; this would probably give a subcritical value of the
threshold, such that the learned sequence is unstable.  This approach
to nonlinear modeling suggests the exciting prospect of designing
models which come close to topological equivalence with the physical
system.

\vskip 2.0pc

\section{Acknowledgments}

We would like to express our gratitude to the {\sl Comit\'e de
Superc\'omputo} of the UNAM, for access to the Cray YM-P4, and other
computer facilities.  One of us (HW) is also indebted to C. Duqu\'e,
director of the {\sl Association pour le D\'eveloppement par la
Recherche et l'Action Integr\'ees}, and the Belgian Ambassador to
Mexico, for their support of our project. The first author (FZ) has
benefited from discussions with P. Gonz\'alez Casanova and
computational advise from G. Kr\"otszch.

\vfill
\eject

\section{References}

\noindent \item{[1]} Hopfield J J 1982 {\it Proc. Natl. Acad. Sci.
USA} {\bf 79} 2554; 1984 Proc. Natl. Acad. Sci. USA {\bf 81} 3088

\noindent \item{[2]} Little W A 1974 {\it Math. Biosci.} {\bf 19} 101

\noindent \item{[3]} Hertz J, Krogh A and Palmer R G 1991 {\it
Introduction to the Theory of Neural Computation} (Addison-Wesley
Redwood City, CA) Ch 2

\noindent \item{[4]} Peretto P 1984 {\it Biol. Cybern.} {\bf 50} 51

\noindent \item{[5]} Amit D, Gutfreund H and Sompolinsky H 1985 {\it
Phys. Rev. A} {\bf 32} 1007

\noindent \item{[6]} Amit D, Gutfreund H and Sompolinsky H 1985 {\it
Phys. Rev.  Lett.} {\bf 55} 1530; 1987 {\it Ann. Phys.} {\bf 173} 30

\noindent \item{[7]} Fontanari J F and Koberle R 1988 {\it J. Phys.
(Paris)} {\bf 49} 13

\noindent \item{[8]} Coolen A C C and Ruijgrok Th W 1988 {\it Phys.
Rev. A} {\bf 38} 4253

\noindent \item{[9]} Shiino M 1990 {\it J. Stat. Phys.} {\bf 59}
1051; Nishimori H, Nakamura T and Shiino M 1990 {\it Phys. Rev. A}
{\bf 41} 3346

\noindent \item{[10]} Coolen A C C and Sherrington D 1992 {\it J.
Phys. A: Math. Gen.} {\bf 25} 5493

\noindent \item{[11]} Bernier O 1991 {\it Europhys. Lett.} {\bf 16}
531

\noindent \item{[12]} Nishimori H and Ozeki T 1993 {\it J. Phys.
 A: Math. Gen.} {\bf26} 859

\noindent \item{[13]} Shukla P 1993 {\it J. Stat.Phys.} {\bf71} 705

\noindent \item{[14]} Zertuche F, L\'opez-Pe\~na R and Waelbroeck H
1994 {\it J. Phys. A: Math. Gen.} {\bf27} 1575

\noindent \item{[15]} Peretto P 1988 {\it J.Phys. (Paris)} {\bf
49} 711; Geszti T 1990 {\it Physical Models of Neural Networks}
(World Scientific, Singapore) Ch 4

\noindent \item{[16]} Arfken G 1970 {\it Mathematical Methods for
Physicists} (Academic Press, New York) Ch 10

\noindent \item{[17]} Penna T J P and Oliveira P M C 1989 {\it
J. Phys. A: Math. Gen.} {\bf22} L719

\vfill
\eject

\section{Figure captions}

\noindent $\underline{Figure \ 1}$.  The increase in the storage
capacity of the neural network with the threshold is represented,
in the deterministic limit $T = 0$. The critical capacity with
threshold equal to zero is $\alpha_c = 0.278$.
With a threshold $\eta = 1$ the critical capacity increases to $0.36$,
and at $\eta = 2$ it reaches $\alpha_c = 1.1$

\noindent $\underline{Figure \ 2}$.  The maximum storage capacity is
given as a function of the threshold at $T = 0.4$. For low values of
the threshold the capacity is less than in the deterministic case
$T = 0$, as thermal fluctuations inhibit a clean recall of
the sequence. However, the improvement in the storage capacity with
the threshold is stronger than at zero temperature, and for
thresholds above $\eta = 1.15$ the capacity of the $T = 0.4$
network surpasses that of the deterministic network.

\noindent $\underline{Figure \ 3}$.  The storage capacity is given as a
function of the threshold at \break $T = 0.7$.

\noindent $\underline{Figure \ 4}$. The results of the numerical
simulation of the network is represented.  The average overlap with the
learned sequence is given as a function of the storage capacity
$\alpha = p/N$, for various values of the threshold. A sharp decline
in the overlap is detected near the theoretical value $\alpha_c = 0.28$
for $\eta = 0$. The increase in capacity from $\eta = 0$ to $\eta = 1$
is roughly equal to $0.06$, again close to the theoretical prediction.
For $\eta = 2$ the capacity increases substantially: one has an
accurate recall of the sequence up to $\alpha = 0.6$, and the
overlap with the sequence goes to zero near the theoretical
critical value $\alpha_c = 1.1$ .

\noindent $\underline{Figure \ 5}$.  The numerical simulation with $
N = 1681 $ neurons reveals a critical capacity $ \alpha_{c} \approx
1.1 $, in agreement with the theoretical calculation.

\bye